\documentclass{IEEEtran}

\usepackage{cite}
\usepackage{amsmath,amssymb,amsfonts}
\usepackage{graphicx}
\usepackage{textcomp}
\usepackage{xcolor}
\usepackage{verbatim}
\usepackage[utf8]{inputenc}
\usepackage{svg}
\usepackage{newunicodechar}
\newunicodechar{−}{\ensuremath{-}}
\usepackage{tikz}
\usepackage{subfig}
\usepackage{cite}
\usepackage{float}
\usepackage{booktabs}
\usepackage{comment}
\usepackage{multirow}
\usepackage{makecell}
\usepackage{threeparttable}
\usepackage{pgfplots }
\usepackage[english]{babel}
\usepackage{amsmath}
\usepackage{graphicx,url}

\usepackage{algorithm}
\usepackage{algorithmic}
\usepackage{graphicx}
\usepackage{stfloats}
\usepackage{xcolor}

\usepackage{subfig} 

\usepackage{amsthm}

\theoremstyle{remark}

\ifCLASSINFOpdf
\else
\fi

\IEEEoverridecommandlockouts
\begin{document}

\title{Multi-Agent Reinforcement Learning for Online Traffic Scheduling in Time-Sensitive Application}

\author{\IEEEauthorblockN{
Marcos Carvalho\IEEEauthorrefmark{1}\IEEEauthorrefmark{3},
Fatih Temiz\IEEEauthorrefmark{3},
Shavbo Salehi\IEEEauthorrefmark{3}
Melike Erol-Kantarci, Fellow, IEEE\IEEEauthorrefmark{3},
Daniel F. Macedo\IEEEauthorrefmark{1}
}
\\
\IEEEauthorblockA{\IEEEauthorrefmark{1}Universidade Federal de Minas Gerais, Brazil\\\IEEEauthorrefmark{3}School of Electrical Engineering and Computer Science, University of Ottawa, Ottawa, Canada \\
E-mails:\{marcoscarvalho, damacedo\}@dcc.ufmg.br}, \{ftemi033, ssale038, melike.erolkantarci\}@uottawa.ca\\
}

\maketitle
\thispagestyle{empty}
\begin{abstract}
Time-sensitive networking (TSN) is increasingly integrated into mobile edge computing (MEC) to support applications with stringent latency requirements, such as extended reality (XR). However, existing TSN scheduling solutions predominantly rely on static optimization techniques or centralized learning models that are based on fixed traffic patterns, limiting their effectiveness in dynamic environments. In practice, MEC environments often host multiple co-located XR traffic flows whose characteristics evolve over time, creating complex inter-queue dependencies that current schedulers fail to capture. Addressing these challenges requires adaptive, decentralized scheduling mechanisms capable of coordinating multiple TSN queues under varying traffic conditions. To this end, this paper proposes a multi-agent reinforcement learning (MARL) framework for TSN scheduling, where each TSN queue is modeled as an autonomous agent. The Heterogeneous-Agent Proximal Policy Optimization (HAPPO) algorithm is employed to explicitly model inter-agent dependencies and jointly optimize service delivery across queues. The simulation results demonstrate that the proposed approach reduces average frame waiting times by up to 26.8\% and worst-case delays by approximately 16.8\%, highlighting its effectiveness in dynamic XR-driven MEC scenarios.
\end{abstract}


\IEEEpeerreviewmaketitle

\section{Introduction}
\label{sec:introdution}
Time-sensitive networking (TSN) has emerged as a key technology for meeting the ultra-low-latency communication requirements of time-sensitive applications. TSN extends standard Ethernet by providing mechanisms such as IEEE 802.1Qbv~\cite{qbv}, which defines a time-aware shaper that manages multiple queues at the egress ports of network devices. Recent studies have investigated the integration of TSN with mobile edge computing (MEC) environments for enhancing the capability of MEC in delivering timely services~\cite{carvalho2025performance}. Many prior works focus on static industrial scenarios, where the number of flows is predefined, traffic is periodic, and packet characteristics such as size and inter-arrival time remain constant over time. Consequently, schedulers often rely on static optimization methods, such as satisfiability modulo theories and integer linear programming~\cite{stuber2023survey}. While effective in static contexts, these methods face significant scalability limitations due to their long adaptation time to network changes~\cite{zhang2024time}. Current advances in reinforcement learning (RL) have improved TSN scheduling in more complex scenarios. For example,~\cite{min2023reinforcement} proposes a categorical deep Q-network (DQN) to identify load-balanced routes by selecting paths that maximize scheduling success. Similarly,~\cite{he2023deepscheduler} introduces an RL-based scheduler using a graph neural network encoder to handle arbitrary topologies and capture dependencies between flows and link states. Likewise,~\cite{islam2024ai} proposes a dynamic scheduling framework combining graph convolutional networks and deep deterministic policy gradient (DDPG) to schedule flows while meeting deadlines and improving admission rates adaptively. The work in ~\cite{dou2025deterministic} proposes the adaptive TSN (A-TSN) framework, which leverages a DQN to optimize time-slot allocation for periodic flows by selecting intervals that minimize latency based on bandwidth usage. Similarly,~\cite{roberty2024configuring} applies proximal policy optimization (PPO) to optimize gate timing across different queues that minimize the end-to-end latency of frames. The study in ~\cite{zhou2021mitigation} proposes using DDPG to adapt time slot durations for different queues across egress ports, mitigating frames that fail to transmit in their scheduled slots due to external factors such as time-synchronization errors (i.e., misbehavior). 

Existing RL-based TSN schedulers primarily rely on centralized single-agent approaches, where a single policy controls all queues jointly. Such formulations do not explicitly model the interactions and coordination among queues serving applications with different quality of service requirements. In addition, existing TSN-RL studies primarily focus on conventional network applications and traffic models~\cite{zhou2026convergence}, while overlooking emerging XR-oriented services, which introduce significantly stricter latency requirements. To address these limitations, we investigate a scenario involving multiple XR applications with time-varying traffic and propose a novel multi-agent reinforcement learning (MARL)-based time-slot assignment scheme. In the proposed approach, each TSN queue is modeled as an agent associated with a specific XR application. Coordination among agents is achieved through centralized training, enabling cooperative decision-making and adaptive online scheduling. This paper adopts Heterogeneous-Agent Proximal Policy Optimization (HAPPO)~\cite{kuba2021trust}, as it allows each agent to specialize while maintaining coordinated control. To show the performance of the MARL method, we compared our approach with single-agent methods and non-learning-based baselines to evaluate the performance of the cooperative multi-agent scheduling. Our results show that HAPPO significantly outperforms single-agent methods, reducing average frame waiting times by 26.8\% and worst-case delays by 16.8\%, while providing more balanced scheduling than heuristic baselines, resulting in reduced tail latencies for lower-load flows.
 



\section{System Model and Problem Formulation}
\label{sec:system_model}

\subsection{System Architecture Overview}
\begin{figure}[htb!]
	\centering
    \includegraphics[width=0.7\columnwidth]{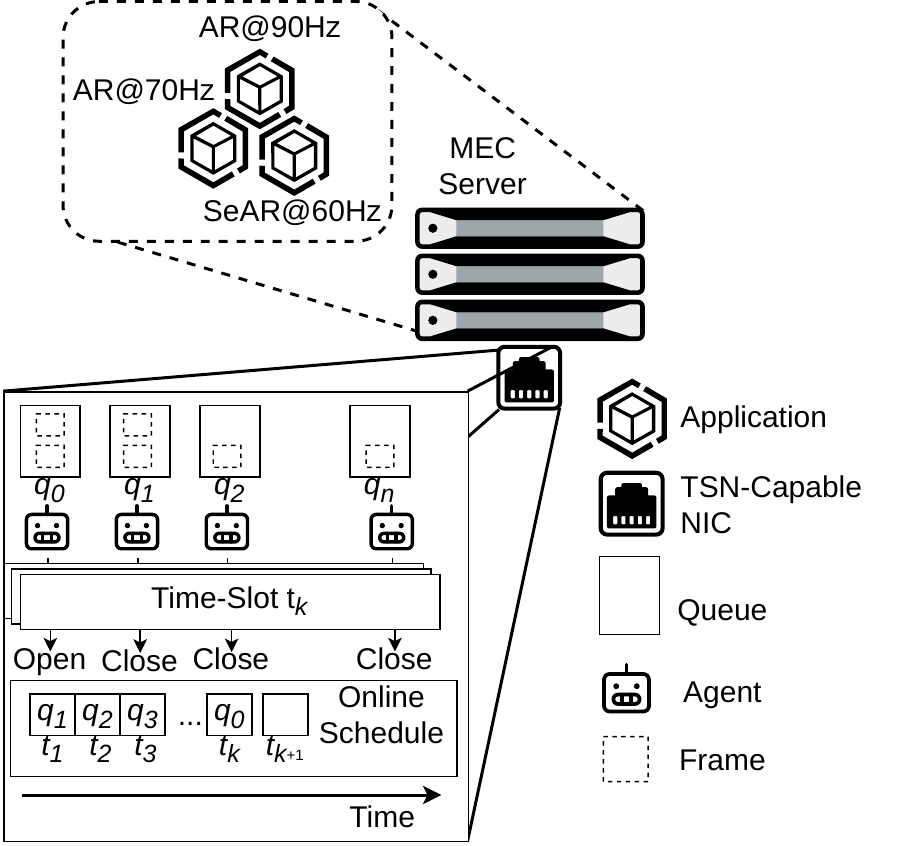}
    \caption{Overall Multi-Agent Online TSN Scheduling}    
    \label{fig:proposedArchitecture}
\end{figure}

In this work, we model the integration of TSN with a MEC server equipped with a TSN-capable network interface card (NIC)~\cite{carvalho2025performance} as illustrated in Fig.~\ref{fig:proposedArchitecture}. The server hosts co-located XR applications with heterogeneous requirements, ranging from full high-resolution video transmission in traditional AR to Semantic AR (SeAR). The latter employs edge-based segmentation to generate binary masks of relevant objects, which are then transmitted as small packets to the end device~\cite{morin2023extended}.


The TSN-capable NIC comprises a set of queues, denoted by $Q = \{{q_0, q_1, \ldots, q_q}\}$, where each queue $q_i \in Q$ is associated with a specific traffic class. In this work, queue $q_{0}$ and queue $q_{1}$ are assigned to distinct AR applications, while queue $q_{2}$ is reserved for the SeAR application. 
Moreover, frames within each queue are managed and transmitted in the downlink (DL) direction following a first-in, first-out policy. Each queue is managed by an autonomous agent, and at each time slot, the agents decide which queue is granted transmission access. 
Unlike offline scheduling methods, this cooperative process enables online scheduling, where the next time slot is dynamically determined based on the current state of all queues.~\looseness=-1

\subsection{Traffic Model for Emerging Time-Sensitive Applications}
The dynamic nature of XR traffic necessitates traffic models that differ from those used for conventional TSN applications, as traffic patterns may evolve over time in response to application behavior and network conditions. Accordingly, the frame arrival time is modeled by the equation below:
\begin{align}
t_a(f_{j}^{(q_{i})}) &= t_{\text{start}}^{(q_{i})} 
                    + (j - 1) T^{(q_{i})} 
                    + \Delta_{j}^{(q_{i})}, \nonumber
\label{eq:arrival_time_queue}
\end{align}

\noindent where $t_a(f_{j}^{(q_{i})})$ denotes the arrival time of the $j$-th frame in queue $q_{i}$. The term $t_{\text{start}}^{(q_{i})}$ represents the start time of the first enqueued frame. $ T^{(q)}$ is the inter-frame interval determined by the application's data generation rate in queue $q_{i}$, and $\Delta_{j}^{(q_{i})}$ captures timing variations introduced by external factors such as application processing delays, resource contention and fragmentation~\cite{shirmarz2024pixels}.~\looseness=-1

\subsection{TSN Schedule and Latency Model}

The scheduling process operates over a sequence of consecutive time slots $\mathcal{T} = \{t_1, t_2, \ldots, t_N\}$, where each time slot has a fixed duration \(D\). At each time slot, exactly one queue is selected for transmission over the DL communication link. A schedule \(s_i \in S\) is defined as an ordered sequence of queue-selection decisions:
\vspace{-0.2cm}
\begin{equation}
s_i = \big(a_i(t_1), a_i(t_2), \ldots, a_i(t_N)\big),
\label{eq:schedule}
\end{equation}
where $a_i(t_k) \in {Q}$ denotes the queue selected for transmission during time slot \(t_k\). This formulation models the scheduling task as a sequential decision-making process, where a scheduling action is performed at every time slot based on the current network conditions. At time slot $t_k$, each queue $q_i$ is represented by a feature vector capturing its current state as presented by the equation below: 
\vspace{-0.2cm}
\begin{equation} 
\textbf{v}_{q_{i},t_{k}} = \langle b_{i,k},\; w_{i,k},\; g_{i,k}\rangle,
\label{eq:queue_parameters}
\end{equation}

\noindent where $b_{i,k}$ is the backlog size, $w_{i,k}$ is the average waiting time of enqueued frames, and $g_{i,k}$ is the age of the oldest frame (AOF) in the queue. The waiting time of the $j$-th frame in queue $q_i$, measures how long the frame waits for an opportunity to be transmitted, 
 calculated using the following equation:
\vspace{-0.2cm}
\begin{equation}
W_i(f_j^{(q_{i})}) = t_s(f_j^{(q_{i})}) - t_a(f_j^{(q_{i})}),
\label{eq:waiting_time}
\end{equation}

\noindent where $t_s(f_j^{(q_{i})})$ is the service start time and $t_a(f_j^{(q_{i})})$ is the arrival time. Based on the waiting time of each frame, we can determine the average waiting time across all frames, as defined in the equation below:
\vspace{-0.2cm}
\begin{equation}
\bar{W}_i = \frac{1}{X_i} \sum_{j=1}^{X_i} W_i(f_j^{(q_{i})}),
\label{eq:average_waiting_time}
\end{equation}
\noindent where $X_i$ is the total number of frames in queue $q_i$ during $t_k$. 

While average waiting time provides a general measure of queueing delay, it does not capture worst-case behavior. To address this limitation, we define the AOF ($g_{i,k}$) of queue $q_i$ at time-slot $t_k$ as the maximum waiting time among all frames currently buffered in the queue, as formalized below:

\vspace{-0.2cm}
\begin{equation}
g_{i,k} = \max_{f_j \in q_i} W_i(f_j^{(q_{i})}).
\label{eq:oldest_frame}
\end{equation}

\subsection{Problem Formulation}
In the context of XR applications, the network must ensure the timely delivery of frames, meaning that the AOF in the queue remains within acceptable limits. Consequently, the goal is to find the optimal schedule $\mathbf{s}_i^*$ that minimizes the AOF, as described by the equation below.

\begin{subequations}\label{eq:objective_TSNScheduling}
\begin{align}
\mathbf{s}^* &= \arg\min_{\mathbf{s} \in S}
\sum_{k=1}^{N} \sum_{q_i \in \mathcal{Q}} g_{i,k}
\label{eq:objective_TSNScheduling_obj}
\\
\text{s.t.} \quad
& \sum_{j=1}^{M} \mathbb{I}_{a_i(t_k)=q_j} \le 1,
\quad \forall k \in \{1,\dots,N\}
\label{eq:C1_TSNScheduling}
\end{align}
\end{subequations}

\noindent The objective aggregates the AOF over all queues and scheduling time slots, capturing the temporal evolution of queueing delays throughout the scheduling horizon. The constraint in eq. (6b) guarantees mutual exclusion, ensuring that only one queue is scheduled for service during each time slot $t_k$, where $\mathbb{I}$ denotes the indicator function, which equals one if the condition is true and zero otherwise.

\section{HAPPO-based Online TSN Scheduling in Dynamic MEC Environments}
\label{sec:proposal}

\subsection{POMDP Formulation}
In this work, the problem is formulated as a multi-agent partially observable Markov decision process (POMDP), defined by the tuple $\langle \mathcal{M}, \mathcal{O}, \mathcal{A}, \mathcal{P}, r, \gamma \rangle$. Here, $\mathcal{M}$ denotes the set of queue agents, each making decisions to cooperatively satisfy the latency requirements of co-located XR applications. $\mathcal{O}$ represents the local observation, and $\mathcal{A} = \prod_{i=1}^{M} \mathcal{A}_i$ is the joint action space. $\mathcal{P} : \mathcal{S} \times \mathcal{A} \times \mathcal{S} \rightarrow [0,1]$ is the transition probability function, $r$ is the reward function, and $\gamma$ is the discount factor balancing immediate and long-term rewards. Below, we describe the three essential components of each agent $m \in \mathcal{M}$.~\looseness=-1

\begin{itemize}
    \item \textit{Observation}: the local observation of each agent at the time-slot $t_{k}$ is defined in Equation~\ref{eq:queue_parameters}.
    \item \textit{Action}: each agent $m \in M$ outputs a scalar score $a^m_{t_{k}} \in [0,1]$ representing its preference for opening its queue at time-slot $t_{k}$. The final schedule selects the agent with the highest score.
    \item \textit{Reward function}: the reward function is modeled as a team reward, defined by \(  r_{t} = \min(R_i)\), where $R_i$ is a vector containing the individual rewards of all agents. The individual reward for each agent $i$ is computed by the equation below:
 
\begin{equation}
R_i =
\begin{cases}
- \alpha \, \lambda_i, & \text{if } \lambda_i > 1,\\[1ex]
- \lambda_i, & \text{otherwise},
\end{cases}
\quad
\label{eq:reward}
\end{equation}

\noindent where $\lambda_i$ is the delay ratio, calculated as the AOF in queue $q_{i}$ divided by the deadline of the application in that queue.

\end{itemize}

The agents aim to maximize the expected cumulative discounted reward, defined as $r_t = \sum_{t=1}^{T} \gamma^{t-1} r_t$, where $\gamma \in [0,1)$ is the discount factor that determines the relative importance of future rewards. Their actions are selected according to a policy function $\pi : s_t \rightarrow P(a_t)$, where $P(a_t)$ denotes the probability distribution over the action space. In our context, each agent learns to jointly optimize the TSN schedule over time, aiming to minimize the AOF in its queue relative to the application's deadline. However, solving this problem in a multi-agent setting is challenging due to the strong interdependence among queues. Since they share the same transmission medium, the action of one agent directly affects the performance of others, making each agent's optimal policy dependent on the joint behavior and increasing the complexity of learning coordinated policies.

\subsection{HAPPO-Based Solution}
To solve the above multi-agent POMDP, we adopt the HAPPO~\cite{kuba2021trust} algorithm, a MARL approach designed for cooperative multi-agent environments. Unlike centralized single-agent RL methods, where a single policy jointly controls all queues, HAPPO enables each agent to learn an independent policy while still coordinating with the remaining agents. This design improves scalability and allows agents to specialize in distinct scheduling behaviors according to their local TSN queue conditions. Each agent maintains an actor network that selects actions based on local observations. A global critic evaluates the joint observations and actions to provide a coordinated learning signal, as presented in Alg~\ref{alg:happo_tsn}. 
\begin{algorithm}[htb!]
\caption{HAPPO-based TSN Scheduling}
\label{alg:happo_tsn}
\small
\begin{algorithmic}[1]
\STATE Initialize actor networks $\{\theta^i\}_{i=1}^{n}$ and global critic $\phi$
\FOR{$k=0,\dots,K-1$}
    \STATE Collect local queue observations $\{o_t^i\}_{i=1}^{n}$
    \STATE Each agent $m$ outputs a score $a_t^m \in [0,1]$
    \STATE Select the queue associated with the highest score
    \STATE Execute transmission and collect global reward $r_t$
    \STATE Store transitions $(o_t^i,a_t^i,r_t,o_{t+1}^i)$
    \STATE Draw a random permutation $\sigma(1:M)$ of agents
    \FOR{$i=1,\dots,M$}
        \STATE Update actor $\theta^{\sigma(m)}$ using PPO-Clip
    \ENDFOR
    \STATE Update global critic $\phi$
\ENDFOR
\end{algorithmic}
\end{algorithm}

Moreover, HAPPO updates agents sequentially using a random permutation instead of simultaneously. This reduces non-stationarity during training, improving coordination and enabling more stable latency-aware scheduling decisions.

\section{Performance Evaluation}
\label{sec:performance_evaluation}

\subsection{Simulation Settings}
We consider a TSN-MEC environment with three TSN queues. Queues $q_0$ and $q_1$ serve immersive video streams at 3840×1920 resolution with refresh rates of 90 Hz and 72 Hz, respectively, each supporting up to four users. Queue $q_2$ supports a SeAR application for 15 users at 1920×720 and 60 Hz, transmitting semantic information instead of full video streams. The inter-frame interval, inter-packet intervals, packet size, and deadlines of all applications for DL direction were derived from the work available in~\cite{morin2023extended} using a Johnson SU distribution. To reduce the action space and accelerate RL, we fixed the TSN time-slot duration ($D$) at 500~$\mu$s, which is sufficient to transmit multiple frames over a 1 Gbps link. The main RL parameters were set as follows: number of agents = 3, episode length = 150, hidden layer size = 64, number of layers = 2, learning rate = $1\times10^{-4}$, and discount factor $\gamma = 0.99$. We adopted values that are widely used in the literature to ensure stable training for all methods. We defined the additional penalty $\alpha$ (eq.~\ref{eq:reward}) as 1.5. To ensure a dynamic environment, the number of active users is randomly selected at the beginning of each episode. For evaluation, a total of 4800 steps were executed, and all results are reported as average and standard deviation across five different runs.

\textbf{Baselines}: HAPPO is evaluated against centralized single-agent PPO and A2C~\cite{mnih2016asynchronous} baselines to assess the benefits of multi-agent coordination while controlling for architectural differences. In addition, we include non-learning heuristics, namely backlog-aware and AOF-aware policies, to provide a benchmark against domain-specific rule-based schedulers. For the backlog-aware heuristic, the queue with the largest backlog at time slot $t_k$ is selected for service. In contrast, for the AOF-aware policy, the queue containing the oldest frame is selected for service.

\subsection{Simulation Results}
The performance results are presented in terms of RL convergence, average waiting time, and the reaction delay (RD) metric defined as \(
\mathrm{RD}_i
= \mathbb{E}\!\left[
g_i(t_k)
\;\middle|\;
b_i(t_k) > 0 \,\wedge\, q_i(t_k) = \text{open}
\right].
\label{eq:reaction_delay}
\) $\mathrm{RD}_i$ is a temporal metric that shows the expected AOF in queue $q_{i}$, conditioned on the queue being non-empty and the queue $q_{i}$ is open.~\looseness=-1 

\begin{figure}[hbt!]
	\centering
    \includegraphics[width=0.7\columnwidth]{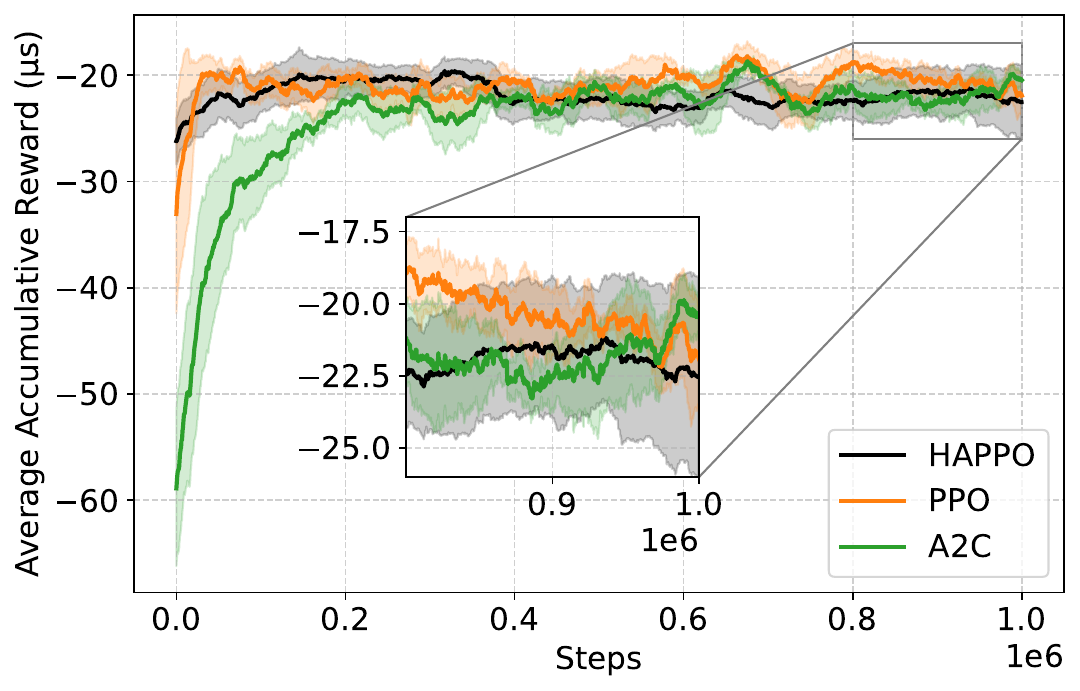}
    \caption{Average Cumulative Reward}    
    \label{fig:training_result}
\end{figure}

\textbf{Learning performance: }Fig.~\ref{fig:training_result} illustrates the training evolution, while Table~\ref{tab:rl_comparison} summarizes the convergence and stability metrics attained during the steady-state phase. 

\noindent The results indicate distinct learning behaviors across the evaluated methods. PPO demonstrated the highest efficiency, reaching initial convergence at approximately 22.2k steps. Furthermore, PPO exhibited superior robustness, with its sensitivity to stochastic initialization being 27.32\% lower than that of A2C. Most notably, PPO reduced temporal volatility (intra-run $\sigma$) by 71.97\% compared to A2C, indicating a significantly more stable policy during steady-state execution. On the other hand, HAPPO required significantly more interactions than PPO to reach initial convergence, stabilizing at approximately 122.9k steps. Furthermore, HAPPO demonstrated the highest sensitivity to stochastic initialization among the evaluated methods, with an inter-seed variability 28.41\% higher than that of A2C and 76.69\% higher than PPO. 
\begin{table}[htb!]
\centering
\caption{Convergence and stability metrics.}
\resizebox{\columnwidth}{!}{
\begin{tabular}{lcccc}
\hline
\textbf{Algo.} & \textbf{Steps (k)} & \textbf{Average Reward} & \textbf{Inter-seeds-$\sigma$} & \textbf{Intra-run-$\sigma$} \\ \hline
PPO            & \textbf{22.2}      & \textbf{-20.77} & \textbf{1.33}           & \textbf{0.95}           \\
HAPPO          & 122.9              & -22.93          & 2.35                    & 0.98                    \\
A2C            & 210.6              & -23.33          & 1.83                    & 3.39                    \\ \hline
\end{tabular}%
}
\label{tab:rl_comparison}
\end{table}

\noindent This suggests that while HAPPO achieves a stable intra-run policy, its performance is highly dependent on the initial seed and by the randomization of the agent update order. Finally, while all methods converged, PPO achieved the highest average reward post-convergence (-20.77), whereas A2C exhibited the lowest terminal performance (-20.45). This gap suggests that by optimizing a single objective without the coordination overhead of mechanisms like HAPPO, agents may avoid conservative updates or reward trade-offs, leading to superior learning performance.~\looseness=-1

\begin{figure*}[htb!]
  \centering
  \subfloat[Queue $q_{0}$]{%
    \includegraphics[width=0.32\linewidth]{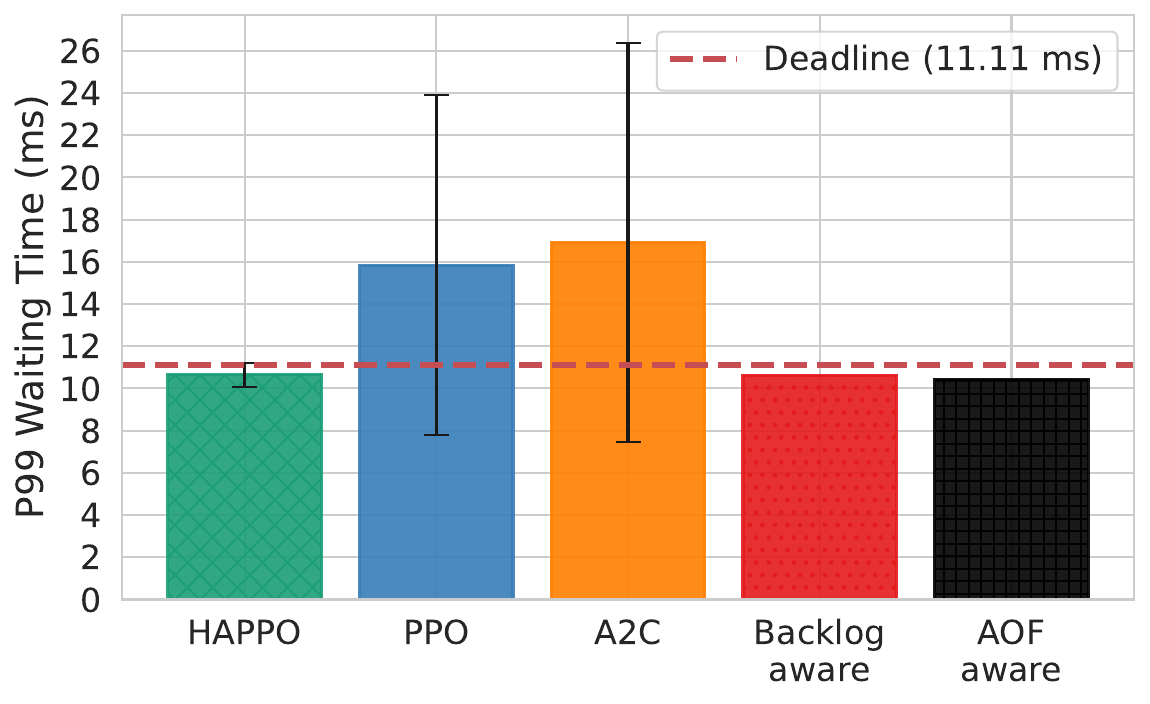}%
    \label{fig:queue0_avg_waiting_time}%
  }
  \hspace{0.1cm}
  \subfloat[Queue $q_{1}$]{%
    \includegraphics[width=0.32\linewidth]{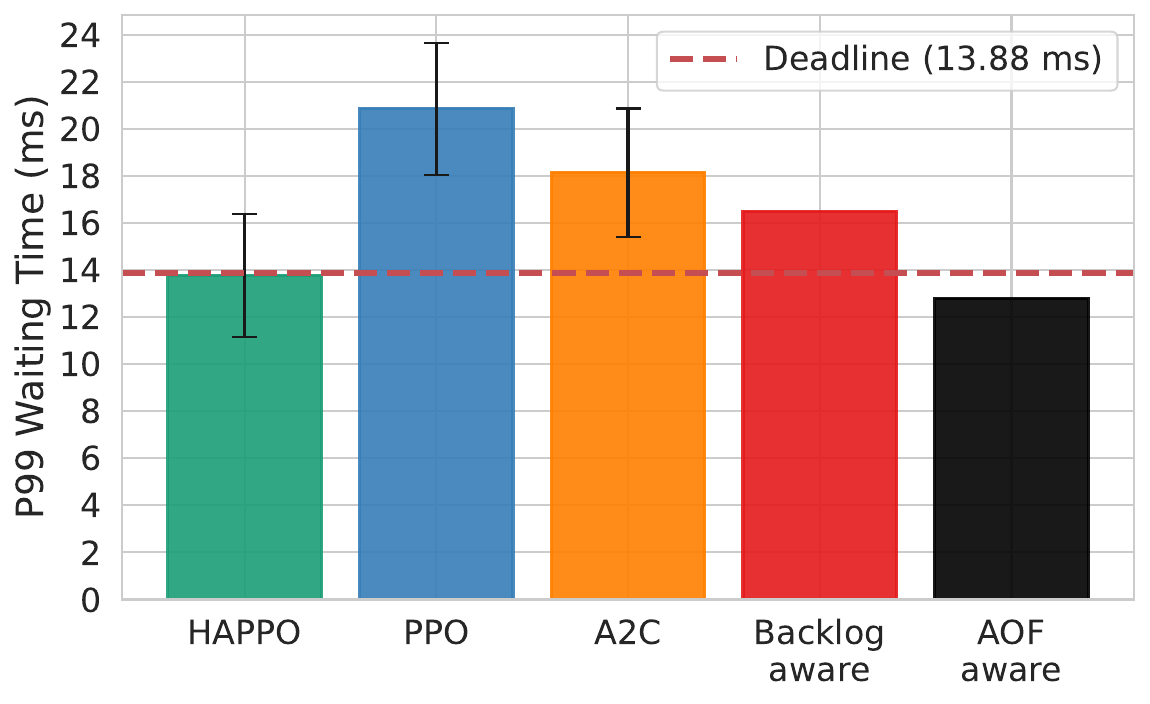}%
    \label{fig:queue_avg_waiting_time}%
  }
  \hspace{0.1cm}
  \subfloat[Queue $q_{2}$]{%
    \includegraphics[width=0.32\linewidth]{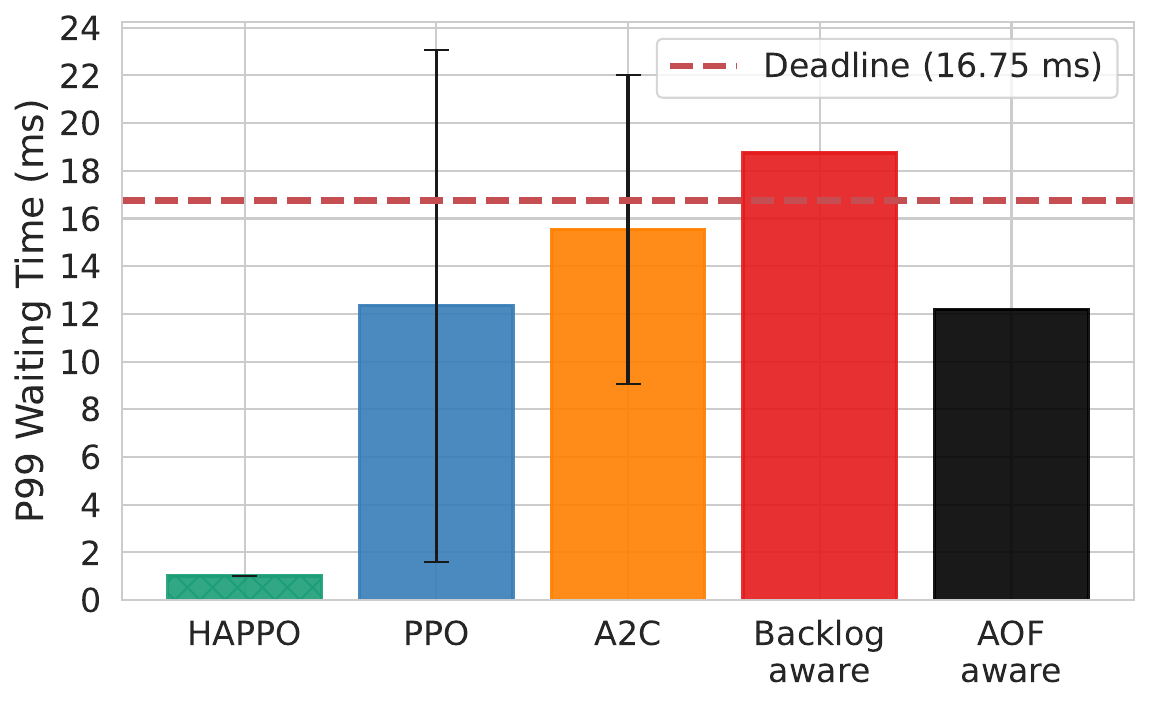}%
    \label{fig:queue2_avg_waiting_time}%
  }

\caption{\textit{p99} of average waiting time}

  \label{fig:queues_eval_avg}
\end{figure*}

\textbf{Network-centric performance: }Although PPO converged faster and A2C achieved the highest final reward, the network performance in Fig.~\ref{fig} and Fig.~\ref{fig} reveals a critical trade-off. As shown in Fig.~\ref{fig:queues_eval_avg}, HAPPO outperforms single-agent methods by minimizing the average frame waiting time across all queues at the \textit{p99} percentile. While HAPPO maintains, on average, the average waiting time below the deadlines across all queues, single-agent methods fail to schedule transmissions in a timely manner, exceeding the deadlines in $q_{0}$ and $q_{1}$ about 48.51\% and 41.06\%, respectively. Although PPO and A2C keep the average waiting time below the deadline in $q_{2}$, their higher variance and elevated tail latencies indicate less predictable timing behavior, which negatively impacts overall system performance. This performance disparity stems from the limitations of centralized single-agent methods, which rely on a single monolithic policy mapping the entire state and thus fail to capture the complex inter-dependencies among queues. In contrast, HAPPO’s multi-agent framework decomposes the problem into coordinated agents that better capture these inter-dependencies and balance transmission across all queues. On the other hand, AOF-aware and backlog-aware heuristics prioritize queues based on instantaneous delay spikes or traffic volume, often at the expense of lower-load flows. For instance, while the AOF-aware heuristic keeps average waiting times below the application deadline across all queues, it does so by delaying packets in the lower-load queue $q_{2}$. In contrast, HAPPO leverages a cooperative multi-agent design to prevent starving $q_{2}$ for the sake of the high-resolution streams in $q_{0}$ and $q_{1}$, ensuring more consistent deadline adherence across heterogeneous applications. 

\begin{figure*}[htb!]
  \centering

  \subfloat[Queue $q_{0}$]{%
    \includegraphics[width=0.32\linewidth]{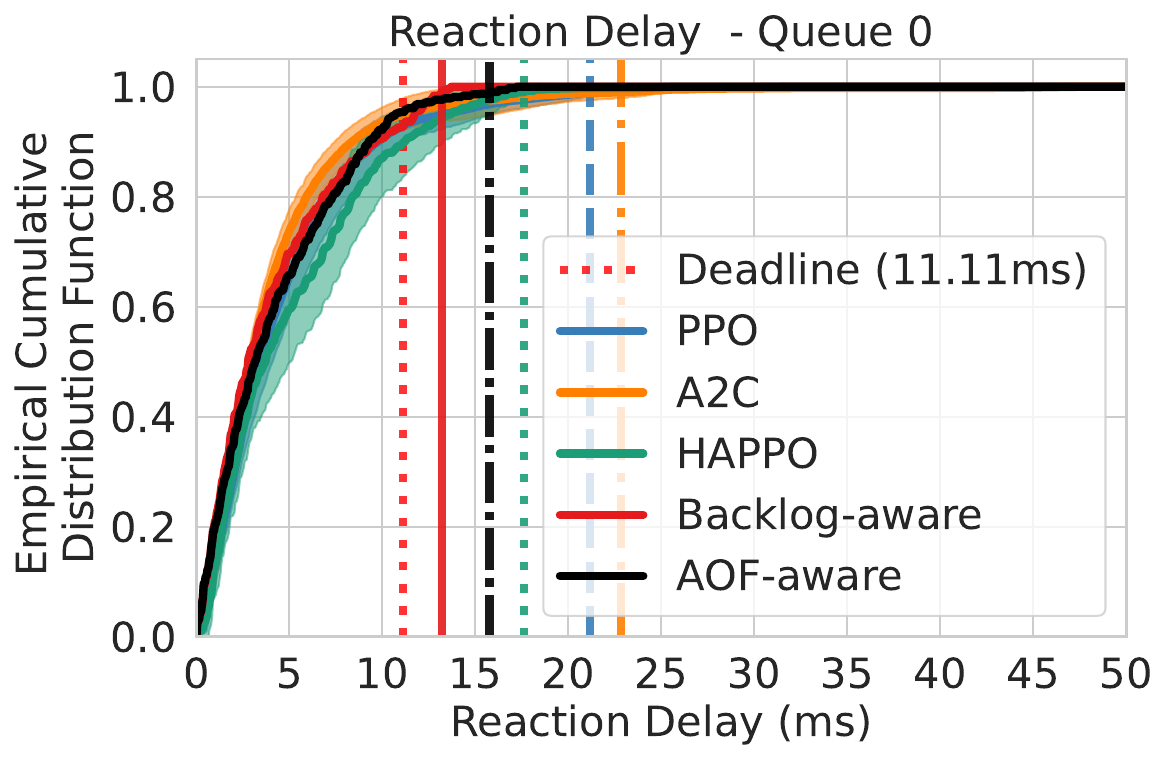}
    \label{fig:queue0_rd}%
  }
  \hspace{0.1cm}
  \subfloat[Queue $q_{1}$]{%
    \includegraphics[width=0.32\linewidth]{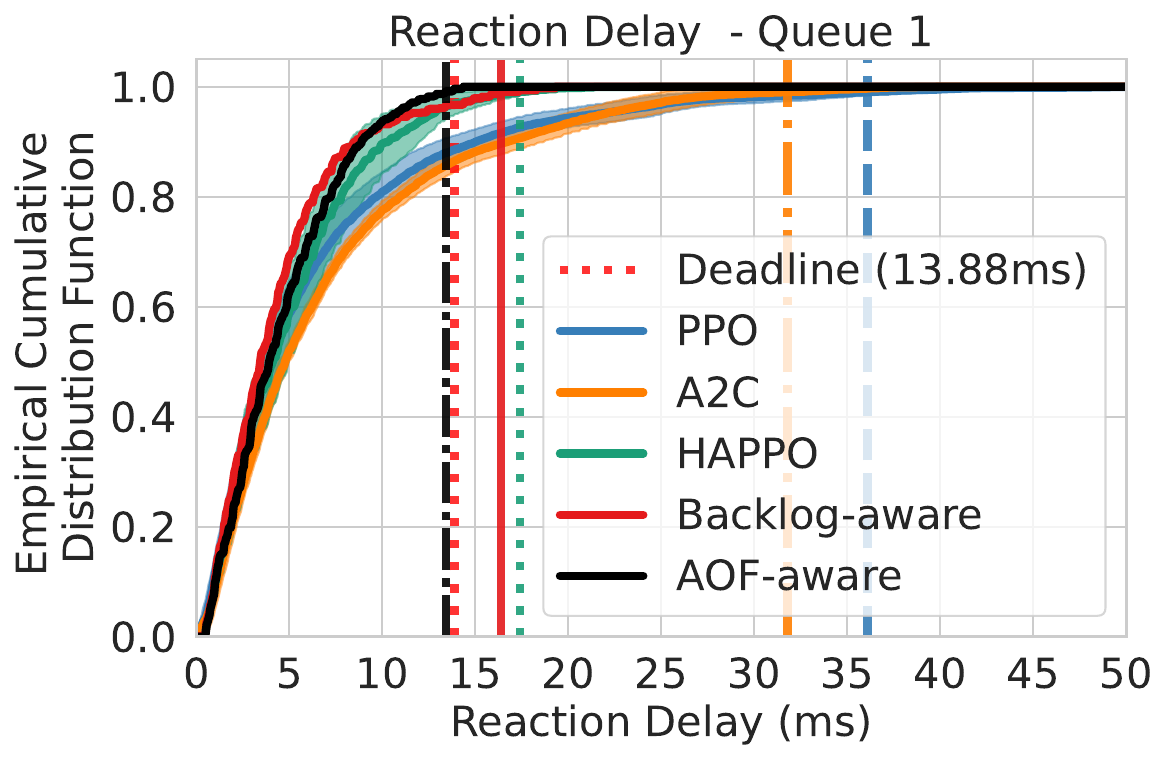}%
    \label{fig:queue1_rd}%
  }
  \hspace{0.1cm}
  \subfloat[Queue $q_{2}$]{%
    \includegraphics[width=0.32\linewidth]{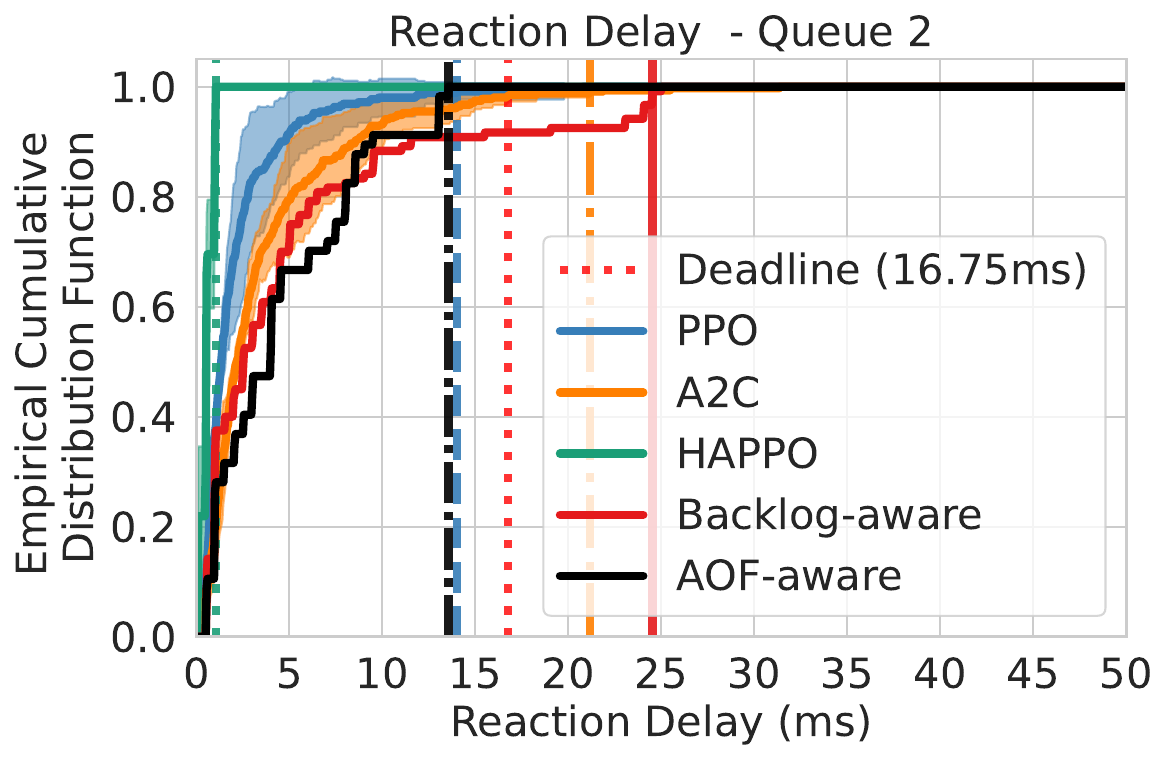}%
    \label{fig:queue_rd}%
  }

\caption{Age of oldest frame distribution. Vertical lines indicate the \textit{p99} value and the application deadlines across the queues.}
  \label{fig:oldest_frame_eval}
\end{figure*}

Fig.~\ref{fig:oldest_frame_eval} depicts the AOF frame when a queue is granted access to transmit (reaction delay). As the results show, at the \textit{p99} percentile, all methods struggle to deliver the last frame on time across all queues. Such missed deadlines were expected because of high workload, network contention, limited link capacity, and the fixed 500~$\mu$s time slot. Nevertheless, HAPPO minimizes these delays by ensuring that the last frame of each flow is delivered as close as possible to its deadline when compared with single agents. For example, in queue $q_{1}$, HAPPO keeps the oldest frame 25.50\% above the deadline, while PPO and A2C exceed it by 160.01\% and 128.96\%, respectively. Moreover, as observed previously, HAPPO demonstrates superior tail-latency management by ensuring that the SeAR traffic in $q_2$ is not sacrificed to mitigate the higher-intensity workloads of the video streams in $q_0$ and $q_1$. This improvement can be attributed to the cooperative coordination among the agents, in contrast to the centralized approach of single agents or the rigid, reactive logic of rule-based heuristics.

\section{Conclusion}
\label{sec:conclusion}
In this paper, we propose a MARL algorithm to improve traffic scheduling across multiple TSN queues for XR applications. To meet the requirements of the XR applications, we utilized the HAPPO algorithm to coordinate services in the MEC environment. We evaluated MARL by comparing it with single-agent methods and rule-based heuristics, considering convergence behavior, average frame waiting time, and worst-case performance based on the AOF. The simulation results show that HAPPO delivers frames within deadlines at high percentiles and minimizes performance degradation when deadlines are missed. In contrast, the baselines struggle to meet deadlines and exhibit higher delays, highlighting the importance of agent cooperation for consistent service. Future work will evaluate the impact of adaptive time-slot duration in the model.

\section*{Acknowledgment}
This work was financed in part by the Coordenação de Aperfeiçoamento de Pessoal de Nível Superior - Brasil (CAPES) - Finance Code 001, CNPq (funding agency from the Brazilian federal government), FAPEMIG (Minas Gerais State Funding Agency), and São Paulo Research Foundation (FAPESP) with Brazilian Internet Steering Committee (CGI.br), grants 2018/23097-3 and 2020/05182-3, NSERC Canada
Research Chairs Program. The authors also gratefully acknowledge the support of the University of Ottawa Visiting Researchers Program/Office.

\bibliographystyle{IEEEtran}
\bibliography{bibtex}

\end{document}